\documentclass[10pt,draft]{dis03}
\usepackage{epsf,amsmath}

\def\lsim{\mathrel{\rlap{\lower4pt\hbox{\hskip1pt$\sim$}}proc.ps
    \raise1pt\hbox{$<$}}}         
\def\gsim{\mathrel{\rlap{\lower4pt\hbox{\hskip1pt$\sim$}}
    \raise1pt\hbox{$>$}}}         

\newlength{\hwlong}
\newcommand{\CP}[1]{
  \settowidth{\hwlong}{CP}%
  CP\unitlength0.7ex%
  \begin{picture}(0,0)
  \put(-4.1,-0.2){\line(5,3){\hwlong\divide\unitlength}}
  \end{picture}%
}

\def\lsim{\mathrel{\rlap{\lower4pt\hbox{\hskip1pt$\sim$}}
    \raise1pt\hbox{$<$}}}         
\def\gsim{\mathrel{\rlap{\lower4pt\hbox{\hskip1pt$\sim$}}
    \raise1pt\hbox{$>$}}}         

\newcommand{\ie}{{\it i.e.}}
\newcommand{\etal}{{\it et al.}}

\newcommand{\GVDM}{\textrm{\tiny GVDM}}

\begin{document}

\title{GENERALISED VDM AND $F_2$ DATA AT LOW $Q^2$
\thanks{Talk by GI at DIS2003 in St.\ Petersburg, to appear in the proceedings.}
}

\author{Johan Alwall and Gunnar Ingelman\\
High Energy Physics, Uppsala University\\ 
Box 535, S-75121 Uppsala, Sweden\\
E-mail: johan.alwall@tsl.uu.se, gunnar.ingelman@tsl.uu.se}

\maketitle

\begin{abstract}
\noindent 
The generalised vector meson dominance model (GVDM) gives a good description of 
$F_2$ data at very low $Q^2$. At intermediate $Q^2$ a GVDM component avoids problems when applying the large-$Q^2$ DIS formalism, such as a negative gluon distribution in the proton. The fluctuations of the exchanged photon into vector mesons is also a natural part of a model with hadronic fluctuations of the target proton, which gives a good description of the non-perturbative $x$-shape of the proton's parton density functions at the starting scale $Q^2_0$ for DGLAP evolution. 
\end{abstract}

\section{Introduction} 
Data on the proton structure function $F_2$ from $ep$ (or $\mu p$) scattering are usually interpreted in terms of the formalism for deep inelastic scattering (DIS) where 
$d\sigma/dxdQ^2 \sim F_2(x,Q^2) = \sum_q e_q^2 \left( xq(x,Q^2) + x\bar{q}(x,Q^2)\right)$.
$F_2$ is, therefore, interpreted in terms of the quark density functions $xq(x,Q^2)$ in the proton and the gluon density enters via the DGLAP equations for evolution in $Q^2$. This formalism has also been applied to $F_2$ data at low $Q^2$ where the exchanged photon is not far from being on-shell. Parametrising such $F_2$ data in terms of quark and gluon density functions results in gluon distributions that tend to be negative at small $Q^2$ \cite{negative-gluon}, since otherwise the strong DGLAP evolution, driven primarily by the gluon at small $x$, gives too large parton densities and thereby a poor fit to $F_2$ in the genuine DIS region at large $Q^2$. Although one may argue that the gluon density is not a directly observable quantity and hence might be negative, it certainly is in conflict with the normal interpretation in terms of the probability for a gluon with momentum fraction $x$ in the proton. In particular, such a gluon distribution may not be universal and applying it in other interactions may, therefore, give incorrect results. 

\begin{figure}[thb]\label{fig:resolution}
\vspace*{45mm}
\begin{center}
\includegraphics{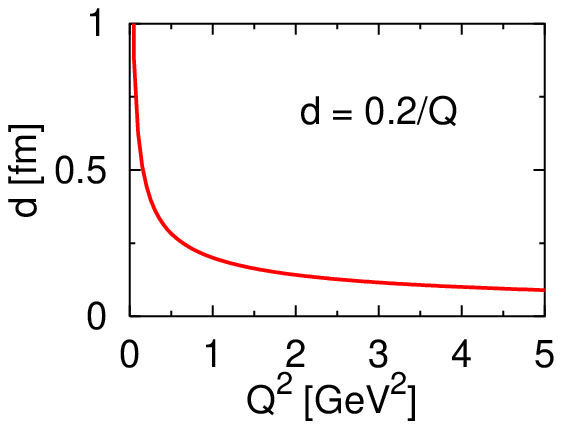}
\includegraphics{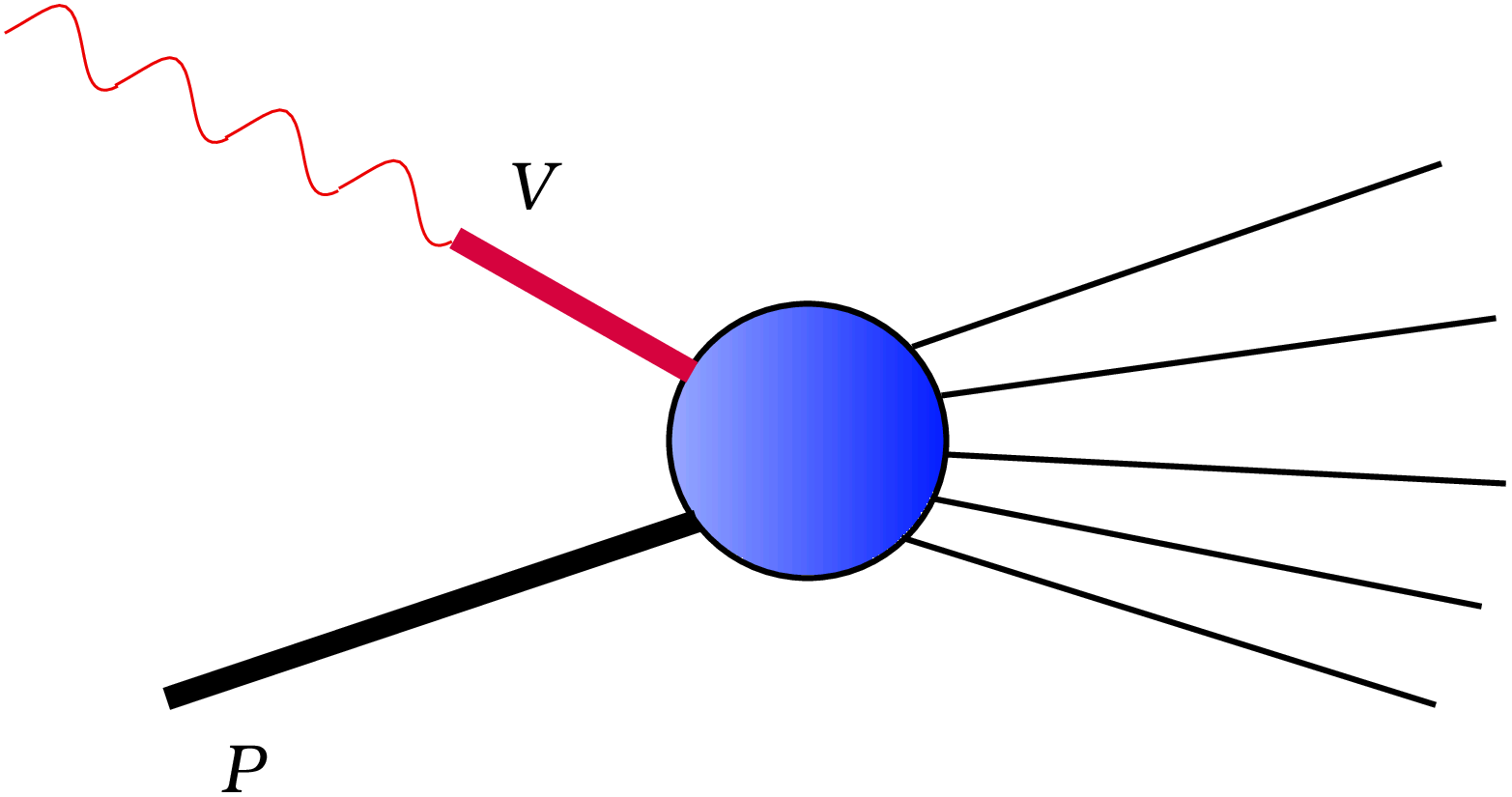}
\vspace*{-16mm}
\caption[*]{(a) Resolved distance scale ($d$ in Fermi) versus $Q^2$ of the probe.\linebreak
(b) Photon fluctuating into a vector meson interacting with the target proton.}
\end{center}
\vspace*{-8mm}
\end{figure}

The mistake is here to apply the formalism for DIS also in the low-$Q^2$ region, where the momentum transfer is not large enough that the parton structure of the proton is clearly resolved, as illustrated in Fig.~\ref{fig:resolution}a. For $Q^2\lsim 1\, \rm{GeV}^2$, there is no hard scale involved and the interaction is a soft photon-proton interaction. The dominant cross-section is here given by the photon fluctuating into a vector meson state which then interacts with the proton in a strong interaction (Fig.~\ref{fig:resolution}b), \ie\ the vector meson dominance model.

\section{Generalised vector meson dominance model in $ep$ at very low $Q^2$}
Due to quantum fluctuations a photon may appear as a vector meson, \ie\  $|\gamma\rangle = C_0|\gamma_0\rangle + \sum_V \frac{e}{f_V}|V\rangle + \int_{m_0}dm_V (\cdots)$. The sum is over the vector meson states $V = \rho^0, \omega, \phi \ldots$ and the integral is for the generalisation to include a continous mass spectrum \cite{GVDM}. This hadronic state then interacts with the target proton, resulting in transverse and longitudinal cross-sections 
$\sigma_{T,L}^{\GVDM}(\gamma p\to X) = \sum_V P_{T,L}(\gamma \to V)\: \sigma_{T,L}(Vp\to X)$, where the fluctuation probability includes the vector meson propagator and is given by GVDM phenomenology \cite{AI}. The soft hadronic cross-section can be taken as the standard parametrisation $\sigma(Vp\to X) = A_V s^\epsilon + B_V s^{-\eta}$, with $\epsilon\approx 0.08$ in the pomeron exchange term which dominates over the reggeon exchange term at high energies. 

In $ep$ scattering, $F_2(x, Q^2) = \frac{(1-x)Q^2}{4\pi^2\alpha} \left( \sigma_T+\sigma_L\right)$ and 
$s_{\gamma p} = Q^2\: \frac{1-x}{x} + m_p^2 \approx Q^2/x$ at small-$x$.
Inserting the full GVDM expressions for $\sigma_{T,L}$ results in \cite{AI} 
\begin{equation}
F_2= \frac{(1-x)Q^2}{4\pi^2\alpha} 
      \left\{  \sum_V r_V \left(\frac{m_V^2}{Q^2 + m_V^2}\right)^2
	\left(1 + \xi\frac{Q^2}{m_V^2}\right) 
     + r_C\frac{m_0^2}{Q^2 + m_0^2}\right\}
	A \frac{Q^{2\epsilon}}{x^\epsilon} \nonumber
\end{equation}
Here, the last factor originating from $\sigma(Vp\to X)$ only includes the pomeron term, since the reggeon term is negligible in the small-$x$ region relevant here. An overall normalisation constant $A$ is introduced giving ratios $A_{V,C}/A$ included in the parameters $r_{V,C}$. In the curly bracket, conventional VDM gives the sum over vector mesons with the characteristic vector meson propagators and the fluctuation constants $r_V=\frac{4\pi\alpha}{f_V^2}\cdot \frac{A_V}{A}$ involving the vector meson decay constant $f_V$. Besides the dominating contribution from transverse photons, the VDM sum contains a longitudinal contribution through the $\xi$-term. The term with $r_C$ ($=1-\sum_V r_V$) originates from the integral over the continous mass spectrum with a lower limit $m_0$ (only the transverse contribution is here included since the longitudinal one is small). Altogether, GVDM gives a more complex $Q^2$ dependence than the simple VDM for transverse photons. The parameters involved are known from GVDM as  
$r_{V=\rho,\omega,\phi,C} = 0.67, 0.062, 0.059, 0.21$, $\xi\approx 0.6$ and $m_0=0.9$ GeV \cite{AI}. 

\section{Comparison to $F_2$ data}
The above expression for $F_2$ compares very well with the HERA $F_2$ data at low $Q^2$, as shown in Fig.~2. The fit gives $\chi^2 = 89/(70-3) = 1.3$ with values as expected for the three free parameters used in the fit, namely $\epsilon = 0.09$, $A = 71\, \mu\rm{b}$, $\xi = 0.6$ \cite{AI}. This demonstrates that at $Q^2$ clearly below 1 GeV$^2$ the HERA $ep$ cross-section can be fully accounted for by GVDM using parameter values as determined from old investigations related to fixed target data. 

\begin{figure}[tbp]\label{fig:lowQ2}
\vspace*{95mm}
\begin{center}
\includegraphics{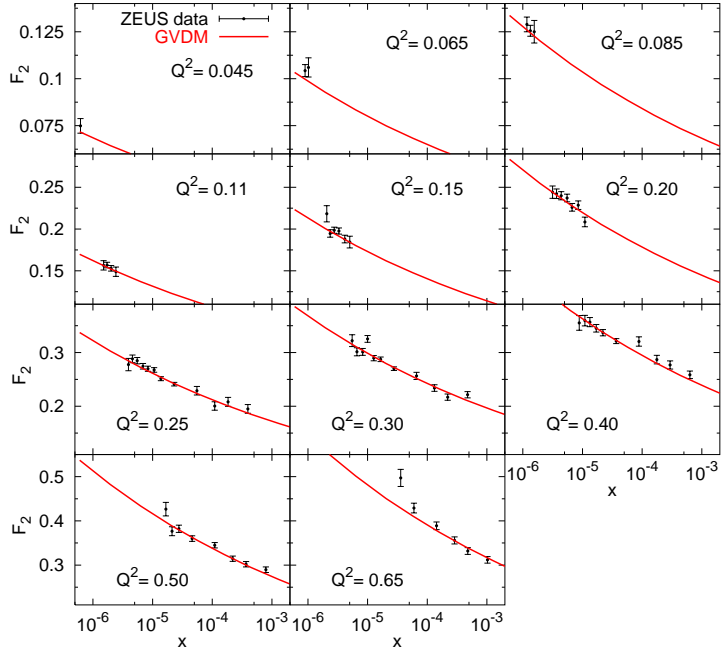}
\vspace*{-15mm}
\caption{$F_2$ at low $Q^2$: GVDM compared to HERA $ep$ data from ZEUS \cite{ZEUS}.}
\end{center}
\vspace*{-5mm}
\end{figure}

\begin{figure}[hbtp]\label{fig:intermediateQ2}
\vspace*{65mm}
\begin{center}
\includegraphics{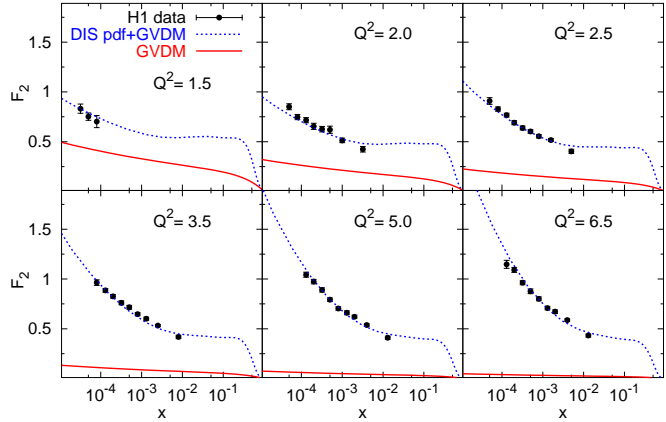}
\vspace*{-15mm}
\caption{$F_2$ at intermediate $Q^2$: GVDM contribution to complete model including DIS parton density functions compared to H1 data \cite{H1}.}
\end{center}
\end{figure}

At larger $Q^2$ GVDM does not give the correct $Q^2$ dependence since the resulting $F_2$ increases with $Q^2$. This may be interpreted physically as a need for a form factor suppression and we introduce the factor $(Q^2_0/Q^2)^a$ for $Q^2>Q^2_0$ to phase out GVDM. Instead, the parton model should become applicable in the DIS region. As shown in Fig.~\ref{fig:intermediateQ2}, a good description of HERA $F_2$ data at intermediate $Q^2$ can be obtained by combining GVDM, with fitted values $a=1.8$ and $Q^2_0=1.26$ in the form factor, and parton density functions that fit HERA $F_2$ data at larger $Q^2$. As can be seen, GVDM gives a negligible contribution for $Q^2\gsim 3\, \rm{GeV}^2$. 

\section{Model for $x$-shape of parton distributions at $Q_0^2$}
The parton distributions used in Fig.~\ref{fig:intermediateQ2} are not just parametrisations, but are obtained from a model \cite{EI} where valence quarks and gluons are derived from momentum fluctuations according to a gaussian distribution with a width given by the uncertainty relation and the proton size. Sea quarks and gluons are obtained from similar momentum fluctuations in hadronic fluctuations of the proton, \ie\   
$|p\rangle  =  \alpha_0|p_0\rangle + \alpha_{p\pi}|p_0\pi^0\rangle + \alpha_{n\pi}|n \pi^+\rangle + \ldots \alpha_{\Lambda K}|\Lambda K^+\rangle + \ldots$
This model gives a good description of available $F_2$ data with only a few fitted parameters \cite{EI}. Furthermore, it gives $u_v(x)\ne d_v(x)$ and $\bar{u}(x)\ne \bar{d}(x)$ in qualitative agreement with data, as well as $s(x)\ne \bar{s}(x)$ of interest for the NuTeV anomaly \cite{AI}. 
This model for parton distributions via hadronic fluctuations, fits very naturally together with GVDM based on hadronic fluctuations of the photon. 

\section{Conclusions} 
The full generalised vector meson dominance model, including contributions from a continous mass spectrum and longitudinal polarisation states, reproduces HERA $F_2$ data at very low $Q^2$ using parameter values in agreement with old analyses of GVDM. Introducing a form factor damping at larger $Q^2$ gives a smooth transition into the deep inelastic region where a description of $F_2$ in terms of parton distribution functions become appropriate.


\begin{thebibliography}{99}
\bibitem{negative-gluon} 
J.\ Pumplin \etal, JHEP 0207:012 (2002)
\bibitem{GVDM} 
J.J.\ Sakurai and D.\ Schildknecht, Phys.\ Lett.\ B40, 121 (1972)
\bibitem{AI} 
J.\ Alwall, G.\ Ingelman, papers in preparation
\bibitem{ZEUS} 
J.\ Breitweg \etal, ZEUS collaboration, Phys.\ Lett.\ B487, 53 (2000)
\bibitem{H1} 
C.\ Adloff \etal, H1 collaboration, Eur.\ Phys.\ J.\ C21, 33 (2001)
\bibitem{EI} 
A.~Edin and G.~Ingelman, Phys.\ Lett.\ B {\bf 432}, 402 (1998)
\end{thebibliography}
\end{document}